\documentstyle[12pt,a4,graphicx]{article}

\newcommand{\be}{\begin{equation}}
\newcommand{\ee}{\end{equation}}
\newcommand{\ba}{\begin{eqnarray}}
\newcommand{\ea}{\end{eqnarray}}
\newcommand{\dis}{\displaystyle}

\begin{document}
\begin{titlepage}
\vspace*{-4cm}
\begin{flushright}
CERN-PH-TH/2006-204\\
UAB-FT-614\\
\end{flushright}
\vspace{1cm}
\begin{center}

{\large\bf $|V_{us}|$  from Strange Hadronic Tau Data
\footnote{
Invited talk given by J.P. 
at ``ICHEP'06, 26-July to 2-August 2006, Moscow, Russia''.}}\\
\vspace{1cm}
{\bf Elvira G\'amiz$^{a)}$, 
Matthias Jamin$^{b)}$,
Antonio Pich$^{c)}$,
 Joaquim Prades$^{d)}$ 
and  Felix Schwab$^{e)}$ 
}\\[0.5cm]
$^{a)}$ Department of Physics, University of Illinois, 
Urbana IL 61801, USA.\\[0.5cm]

$^{b)}$ ICREA and Departament de F\'{\i}sica Te\`orica (IFAE)\\
Universitat Auton\`oma de Barcelona,
 E-08193 Bellaterra, Barcelona, Spain.\\[0.5cm]

$^{c)}$ Departament de F\'{\i}sica Te\`orica, IFIC, 
Universitat de Val\`encia--CSIC \\  Apt. de Correus 22085, E-46071
Val\`encia, Spain.\\[0.5cm]

$^{d)}$ Theory Unit, Physics Department, CERN, CH-1211 Gen\`eve 23,
Switzerland.\\[0.5cm]

$^{e)}$ Physik Department, Technische Universit\"at M\"unchen,
D-85748 Garching and Max Planck Institut f\"ur Physik 
 D-80805 M\"unchen, Germany.\\[0.5cm]
\end{center}
\vfill
\begin{abstract}
\noindent
We report on recent work to
determine the CKM matrix element $|V_{us}|$ using strange hadronic
$\tau$ decay data. We use the recent OPAL update of the strange
spectral function, while on the theory side we update the
dimension-two perturbative contribution including the recently
calculated $\alpha_s^3$ terms. Our result $|V_{us}|=0.2220 \pm
0.0033$ is already competitive with the standard extraction from
$K_{e3}$ decays and other new proposals to determine $|V_{us}|$. The
actual uncertainty on $|V_{us}|$ from $\tau$ data is dominated
largely by experiment and will eventually be much reduced by
B-factories and future $\tau$-charm factory data, providing one of
the most accurate determinations of this Standard Model parameter.
\end{abstract}
\vfill
October 2006
\end{titlepage}

\section{Introduction and Motivation}

BaBar and Belle are just starting to produce their first results on
hadronic tau decays and in particular on Cabibbo-suppressed
modes.\cite{PISA06} These results  will eventually further increase
the high precision status of observables such as
 \ba R_\tau\equiv
\frac{\Gamma[\tau^- \to {\rm hadrons} \, \, \nu_\tau (\gamma)]}
{\Gamma[\tau^- \to e^- \overline \nu_e \nu_\tau (\gamma)]} \, ,\ea
 attained by ALEPH and OPAL at LEP and CLEO at CESR. This
status has been already exploited in a very successful determination
of $a_\tau\equiv \alpha_s(M_\tau)/\pi$.\cite{BNP92,CLEO95,ALEPH98,OPAL99,ALEPH05}
On the other hand, SU(3) breaking effects are sizeable in the
semi-inclusive hadronic $\tau$-decay width into Cabibbo-suppressed
modes.\cite{ALEPH99,CLEO03,OPAL04,CHD97,DHZ06}

These two facts make the strange hadronic $\tau$ decay data an ideal
place for determining SU(3) breaking parameters such as $|V_{us}|$
and/or $m_s$. The obvious advantage of this procedure is that the
experimental uncertainty will eventually be reduced at the
B-factories and at future facilities like the $\tau$-charm factory
BEPCII 
at Beijing.

\section{Theoretical Framework}

Using analytic properties of  two-point correlation functions for
vector ($\mathcal{J}=V$) and axial-vector ($\mathcal{J}=A$) two
quark-currents, \ba \lefteqn{\Pi^{\mu\nu}_{\mathcal{J}, ij}(q)
\equiv i \int {\rm d}^4 x \, e^{i q \cdot x} \langle 0 | T [
\mathcal{J}_{ij}^{\mu\dagger}(x)
\mathcal{J}^\nu_{ij}(0) ]| 0 \rangle}&&  \nonumber \\
&& \equiv \left[q^\mu q^\nu - q^2 g^{\mu\nu}\right]
\Pi^T_{\mathcal{J}, ij}(q^2) + q^\mu q^\nu \Pi^L_{\mathcal{J},
ij}(q^2) \, , \nonumber\\ &&\ea
 one can  express $R_\tau$  as a contour integral
running counter-clockwise around the circle $|s|=M_\tau^2$ in the
complex s-plane:
 \ba R_\tau &\equiv& - i \pi \oint_{|s|=M_\tau^2}
\frac{{\rm d} s}{s} \, \left[1-\frac{s}{M_\tau^2}\right]^3
 \\ &\times& \left\{ 3 \left[1+\frac{s}{M_\tau^2}
\right] D^{L+T}(s) + 4 \, D^L(s) \right\} \, . \nonumber
\ea
We have used integration by parts to rewrite $R_\tau$
in terms of the logarithmic derivatives
\ba
D^{L+T}(s) \equiv - s \, \frac{{\rm d}}{{\rm d } s }
\, \Pi^{L+T}(s) \, ; \nonumber \\
D^{L}(s) \equiv \frac{s}{M_\tau^2} \, \frac{{\rm d} }{{\rm d } s }
\, \left[ s \Pi^{L}(s) \right] \, \, .
\ea
Moreover, one can
experimentally decompose $R_\tau$ into
\be R_\tau \equiv R_{\tau, V}
+  R_{\tau, A} + R_{\tau, S}\, , \ee
according to the quark content
\ba \Pi^J(s)&\equiv& |V_{ud}|^2 \left\{ \Pi^J_{V,ud}(s) +
\Pi^J_{A,ud}(s) \right\}
 \nonumber \\ &+&
|V_{us}|^2 \left\{ \Pi^J_{V,us}(s) + \Pi^J_{A,us}(s) \right\}  ,
\quad \ea
where $R_{\tau, V}$ and $R_{\tau, A}$ correspond to the
first two terms in the first line and $R_{\tau, S}$  to the second
line, respectively.

Additional information can be obtained from the measured
invariant-mass distribution of the final hadrons, which defines the
moments \be \label{OPE} R_\tau^{kl} \equiv {\dis \int^{M_\tau^2}_0}
{\rm d} s \left(1-\frac{s}{M_\tau^2}\right)^k \, \left(
\frac{s}{M_\tau^2} \right)^l \, \frac{{\rm d} R_\tau}{{\rm d} s} \,
. \ee
 At large
enough Euclidean $Q^2=-s$, both $\Pi^{L+T}(Q^2)$ and $\Pi^{L}(Q^2)$
can be organised in a dimensional operator series using well
established QCD operator product expansion (OPE)  techniques. One
gets then
\ba \lefteqn{R_\tau^{kl} \equiv N_c S_{\rm EW} \Big\{ (|V_{ud}|^2 +
|V_{us}|^2) \,  \left[ 1 + \delta^{kl(0)}\right]} &&
 \nonumber \\
 &&\;\mbox{} + {\dis \sum_{D\geq2}} \left[ |V_{ud}|^2 \delta^{kl(D)}_{ud}
+ |V_{us}|^2 \delta^{kl(D)}_{us} \right] \Big\} \, .\quad\; \ea
The electroweak radiative correction $S_{\rm EW}=1.0201\pm0.0003$
has been pulled out\cite{Sew} explicitly and $\delta^{kl(0)}$
denotes the purely perturbative dimension-zero contribution.  The
symbols $\delta^{kl(D)}_{ij}$ stand for  higher dimensional
corrections in the OPE from dimension  $D\geq 2$ operators, which
contain  implicit $1/M_\tau^D$ suppression
factors.\cite{BNP92,PP98,PP99,CK99} The most important being the
operators $m_s^2$ with $D=2$  and $m_s \langle \overline q q
\rangle$ with $D=4$.

 In addition, the flavour  SU(3)-breaking quantity
\ba
\delta R^{kl}_\tau &\equiv&
\frac{R^{kl}_{\tau, V+A}}{|V_{ud}|^2}-
\frac{R^{kl}_{\tau, S}}{|V_{us}|^2} \nonumber \\
&=& N_c \, S_{EW}   \, {\dis \sum_{D\geq 2}} \left[
\delta^{kl(D)}_{ud}-\delta^{kl(D)}_{us}\right]\quad \ea enhances the
sensitivity to the strange quark mass. The dimension-two correction
$\delta^{kl(2)}_{ij}$ is known to order $\alpha_s^3$ for both
correlators, $J=L$ and $J=L+T$.\cite{PP98,PP99,BCK05}

In Ref.~\cite{PP98}, an extensive analysis of this $D=2$
correction was done and it was shown that the perturbative $J=L$
correlator behaves very badly. The $J=L+T$ correlator was also
analysed there to order $\alpha_s^2$ and showed a relatively good
convergence. Here, we have included the recently calculated
$O(\alpha_s^3)$ correction for $J=L+T$.\cite{BCK05} One can see that
the $J=L+T$ series also starts to show its  asymptotic character at
this order, though it is still much better behaved than the $J=L$
component. These series  show clearly an asymptotic behaviour and it
does not make much sense to sum all known orders.

\section{
Determination of $|V_{us}|$}

   One can use the  relation
--and analogous relations for other moments--
\be \label{Vus}
|V_{us}|^2 = \frac{R_{\tau, S}^{00}}{\dis \frac{R_{\tau,
V+A}^{00}}{|V_{ud}|^2}-\delta R_{\tau, {\rm th}}^{00}}\; \ee
to determine the  Cabibbo--Kobayashi--Mas\-ka\-wa  (CKM) matrix
element $|V_{us}|$. Notice that, on the right-hand side of
(\ref{Vus}), the only theoretical input is $\delta R_{\tau, {\rm
th}}^{00}$, which is around 0.25 and gets compared to the experimental
quantity $R_{\tau, V+A}^{00}/|V_{ud}|^2$ which is around 3.7.
Therefore, with a not so precise theoretical prediction for $\delta
R_{\tau, {\rm th}}^{00}$ one can get a quite accurate value for
$|V_{us}|$, depending on the  experimental accuracy.

 The very bad QCD behaviour of the $J=L$ component in
$\delta R_{\tau, {\rm th}}^{kl}$   induces a large  theoretical
uncertainty, which can be reduced considerably using phenomenology
for the scalar and pseudo-scalar
correlators.\cite{GJPPS03,JAM03,GJPPS05}
In particular, the pseudo-scalar spectral functions are dominated by
far by the well-known kaon pole,  to which we add suppressed
contributions from the pion pole, as well as higher excited
pseudo-scalar states whose parameters have been estimated in
Ref.~\cite{MK02}.  For the strange scalar spectral function, we
take the result obtained\cite{JOP00} from a study of S-wave $K\pi$
scattering within resonance chiral perturbation theory,\cite{EGPR89}
which has been recently updated in Ref.~\cite{JOP06}.

 The smallest theoretical uncertainty arises for the
 $kl=00$ moment, for which we get
\ba \label{deltaR} \delta R_{\tau, {\rm th}}^{00}
&=& 0.1544\; (37) + 9.3\; (3.4) \, m_s^2 \nonumber \\
&+& 0.0034\; (28)\; = \; 0.240\; (32) \, , \quad\ea
where $m_s$ denotes the strange quark mass, in GeV units, in the
$\overline{\mathrm{MS}}$ scheme at $\mu=2$ GeV.
 The first term contains the phenomenological scalar and pseudo-scalar
contributions, the second term contains the rest of the perturbative
$D=2$ contribution, while the last term stands for the rest of the
contributions. Notice that the phenomenological contribution is
 more than 64\% of the total, while the rest comes almost from the
perturbative $D=2$ contribution.
 Here, we update $\delta R_{\tau, {\rm th}}^{00}$
in Refs.~\cite{GJPPS03,JAM03,GJPPS05} in various respects. First,
we use the recently updated scalar spectral function\cite{JOP06};
second, we  include the $\alpha_s^3$ corrections to the $J=L+T$
correlator calculated in Ref.~\cite{BCK05} and finally, we use an
average of contour improved\cite{DP92} and fixed order perturbation
results for the asymptotically summed series. A detailed analysis
will be presented elsewhere.\cite{GJPPS}

For the $m_s$ input value, we use the recent average $m_s ( 2\,{\rm
GeV}) = (94 \pm 6)$ MeV,\cite{JOP06} which includes the most recent
determinations of $m_s$ from QCD sum rules and  lattice QCD. The
strange quark mass uncertainty corresponds to the most precise
determination from the lattice.

Recently Maltman and Wolfe criticised the theory error we previously
employed for the $D=2$ OPE coefficient.\cite{MW06} In our updated
estimate (\ref{deltaR}), we have decided to include a more
conservative estimate of unknown higher-order corrections by using
an average of contour improved and fixed-order perturbation theory.
Notice however, that $\delta R_{\tau, {\rm th}}^{00}$ is dominated
by the scalar and pseudo-scalar contributions which are rather well
known from phenomenology, and that the larger perturbative
uncertainty is compensated by the smaller $m_s$ error, so that our
final theoretical uncertainty is almost the same as in previous
works.\cite{GJPPS03,JAM03,GJPPS05}

In order to finally determine $|V_{us}|$, we employ the following
updates of the remaining input parameters: $|V_{ud}|=0.97377 \pm
0.00027$,\cite{PDG06} the non-strange branching fraction $R_{\tau,
V+A}^{00}= 3.471 \pm 0.011$,\cite{DHZ06} as well as the strange branching
fraction\cite{DHZ06} $R_{\tau, S}^{00} = 0.1686 \pm 0.0047$  (see also
Refs.~\cite{ALEPH99} and \cite{OPAL04}), which includes the
theoretical prediction for the decay $B[\tau \to K \nu_\tau
(\gamma)] = 0.715 \pm 0.003$ which is based on the better known
$K\to \mu \nu_\mu(\gamma)$ decay rate. For $|V_{us}|$, we then
obtain
 \be |V_{us}| = 0.2220 \pm 0.0031_{\rm exp} \pm 0.0011_{\rm th} \, . \ee
 The experimental uncertainty includes a small component
from the error in $|V_{ud}|$, but it is dominated by the uncertainty
in $R_{\tau, S}^{00}$, while the theoretical error is dominated by the
uncertainty in the perturbative expansion of the $D=2$ contribution.

\section{Results and Conclusions}

 High precision Cabibbo-suppressed hadronic tau data from
ALEPH and OPAL at LEP and CLEO at CESR provide  already a
competitive result for  $|V_{us}|$. As presented above and in
Refs.~\cite{GJPPS03,JAM03,GJPPS05}, the final uncertainty in the
$\tau$ determination of $|V_{us}|$ becomes an experimental issue and
will eventually be much reduced with the new B-factories
data\cite{PISA06} and further reduced at future $\tau$ facilities. A
combined fit to determine both $|V_{us}|$ and $m_s$ will then be
possible. Hadronic $\tau$ decays have the potential to provide the
most accurate measurement of $|V_{us}|$ and a very competitive $m_s$
determination.

\section*{Acknowledgments}
This work has been supported in part by the European Commission
(EC) RTN FLAVIAnet under
Contract No. MRTN-CT-2006-035482, 
by MEC (Spain) and FEDER (EC) Grant Nos. FPA2005-02211 (M.J.),
FPA2004-00996 (A.P.) and FPA2003-09298-C02-01 (J.P.), by Junta de
Andaluc\'{\i}a Grant Nos. FQM-101 (J.P.) and FQM-437 (E.G. and
J.P.) and  by BMBF (Germany)
Contract Nos. 05HT4WOA/3 and 05HT6WOA (F.S.).

\vfill
\small
{$^*$
On leave of absence from
{\it CAFPE and Departamento de F\'{\i}sica Te\'orica y del
Cosmos, Universidad de Granada, Campus de Fuente Nueva,
E-18002 Granada, Spain.}}

 \end{document}